\documentstyle[11pt]{article}

\newcommand{\be}{\begin{equation}}
\newcommand{\ee}{\end{equation}}
\newcommand{\ii}{{\rm i}}
\newcommand{\eq}{\: = \:}
\newcommand{\id}{{\sf 1 \hspace{-0.3ex} \rule{0.1ex}{1.6ex}
                  \rule{0.3ex}{0.2ex}}}
\newtheorem{note}{Note}

\begin{document}
\begin{center}
{\Large {\bf On a non-CP-violating electric dipole moment of
            elementary particles}} \\
J. Giesen, Institut f\"ur Theoretische Physik, Bunsenstra{\ss}e 9,
D-37073 G\"ottingen \\
(e-mail: giesen@theo-phys.gwdg.de)
\end{center}

\section*{Abstract}
A description of elementary particles should be based on irreducible
representations of the Poincar\'e group. In the theory of massive
representations of the full Poincar\'e group there are essentially
four different cases. One of them corresponds to the ordinary Dirac
theory. The extension of Dirac theory to the remaining three cases
makes it possible to describe an anomalous electric dipole moment of
elementary particles without breaking the reflections.

\section*{Introduction}

For a long time now there has been great experimental and theoretical
interest in an electric dipole moment of elementary particles
\cite{bernreuther,ramsey,salpeter}. We want to characterize an
electric dipole moment of an elementary particle by means of two
properties. First the dipole moment should show the same
behaviour under the reflections (space inversion, time inversion and
charge conjugation) as a classical dipole does. This means that an
electric dipole moment should not be CP-violating. Second an
elementary particle possessing an electric dipole moment should
interact with an external electric field.

A description of elementary particles by the Dirac equation easily
allows the implementation of external fields. This seems to be the
right frame to describe an elementary particle with electric dipole
moment. The ordinary Dirac equation only predicts a magnetic dipole
moment. Nevertheless one has the possibility to add effective terms
to the Lagrangian from which the Dirac equation can be derived. Such
effective terms have been used to model the anomalous values of proton
and neutron magnetic dipole moments \cite{pauli}. Landau \cite{landau}
has shown that it is impossible to find an effective term which allows
the describtion an electric dipole moment. He argued that the spin is
the only relativistic observable which is a vector and can be
constructed in the particles rest frame. An electric dipole moment
which behaves classically should not vanish in the rest frame and
cannot behave like the spin under reflections. Thus Landau denied the
existence of an  electric dipole moment of elementary particles in
the sense stated above.

Nowadays \cite{frere} one no longer demands the first property.
Therefore it becomes possible to describe an anomalous electric dipole
moment by means of the spin. The coupling ${\bf \sigma}{\bf E}$ then
becomes CP-violating.

Wigner \cite{wigner} has pointed out that the irreducible
rep\-resentations of the Poincar\'e group should be the basis for
a description of elementary particles. Therefore we start with
irreducible representations of the Poincar\'e group and not with the
Dirac equation. These representations are not unique when reflections
are taken into ac\-count. A careful examination of the representation
theory in the massive case shows that essentially one has to
distinguish four cases \cite{wigner}. One of these four cases
corresponds to the ordinary Dirac theory. The extension of
Dirac theory to the remaining three cases enables us to describe a
non-CP-violating elementary particle electric dipole moment.

So far we deal with a one particle description. To get a many
particle description one has to quantize the Dirac field. This is
possible with the assumption of weak external fields. One can use
common techniques to construct a Fock space on which one has a
reducible representation of the full Poincar\'e group.

A more detailed discussion of this work can be found in
\cite{giesen}.

\section*{Irreducible representations of Poincar\'e group}

In quantum mechanics one is interested in projective
rep\-resentations of symmetry groups. Bargmann \cite{bargmann} was
able to show that in the case of the restricted Poincar\'e group
every projective representation is totally characterized by an
ordinary representation of the universal covering group. Looking at
the full Poincar\'e group including reflections the situation is a
little bit more complicated. Here one has to take into account that
one is dealing with projective representations. The irreducible
representations for nonvanishing mass have been enumerated by Wigner
\cite{wigner}. He showed that one has to distinguish four cases. In
three of them there is a phenomenon called doubling. The three
doubled representations may be obtained by the usual one by doubling
the number of components. The operators are then written in the form
of a direct product $U(\Lambda ) \eq U_{0} (\Lambda ) \otimes
I(\Lambda )$ of 2$\times$2-matrices $I(\Lambda )$ and operators
$U_{0} (\Lambda )$ which act in the subspace of the representation
of the restricted Poincar\'e group. This subspace consists of
normalizable functions of 2$s+$1 components. The type of a
representation depends on the squares of $U(\Lambda_{{\rm t}} )$ and
$U(\Lambda_{{\rm st}} )$, where $\Lambda_{{\rm t}}$ and
$\Lambda_{{\rm st}}$ denote the time inversion and combined space
and time inversion, respectively. These squares can be normalized to
$\pm \id$. The possible cases are listed in table (1). This means
that time inversion and combined space and time inversion are
represented by antiunitary operators. In the following we only deal
with $s=\frac{1}{2}$ representations. \newpage

\begin{table*}
\begin{tabular}{r||r|r||c|c|c}
    & $U(\Lambda_{{\rm t}} )^{2}$ & $U(\Lambda_{{\rm st}} )^{2}$ &
      $U(\Lambda_{{\rm s}})$ &  $U(\Lambda_{{\rm t}})$ &
      $U(\Lambda_{{\rm st}} )$ \\ \hline \hline
I   & $(-\id )^{2s}$  & $(-\id )^{2s}$  & $\id$ & $C$J & $C$J \\
\hline
II  & $(-\id )^{2s}$  & $-(-\id )^{2s}$ & $\left( \begin{array}{cc}
\id & 0 \\ 0 & -\id \end{array} \right)$ & $\left( \begin{array}{cc}
0 & C \\ C & 0 \end{array} \right)$J & $\left( \begin{array}{cc} 0 &
C \\ -C & 0 \end{array} \right)$J \\ \hline
III & $-(-\id )^{2s}$ & $(-\id )^{2s}$  & $\left( \begin{array}{cc}
\id & 0 \\ 0 & -\id \end{array} \right)$ & $\left( \begin{array}{cc}
0 & C \\ -C & 0 \end{array} \right)$J & $\left( \begin{array}{cc} 0 &
C \\ C & 0 \end{array} \right)$J \\ \hline
IV  & $-(-\id )^{2s}$ & $-(-\id )^{2s}$ & $\left( \begin{array}{cc}
\id & 0 \\ 0 & \id \end{array} \right)$ & $\left( \begin{array}{cc}
0 & C \\ -C & 0 \end{array} \right)$J & $\left( \begin{array}{cc} 0
& C \\ -C & 0 \end{array} \right)$J \\
\end{tabular}
\caption[]{Representation of the reflections \footnotemark[1]
           \footnotemark[2]}
\end{table*}

\footnotetext[1]{$C$ is a representation of $\left(
\begin{array}{cc} 0 & -1 \\ 1 & 0 \end{array} \right) \in $SU(2) in
a space of dimension 2$s +$1.}
\footnotetext[2]{J is the operator of complex conjugation.}

\section*{Electric dipole moments}

It is possible to show that the ordinary Dirac equation
\be \label{d1}
 (\ii \gamma^{\mu} \partial_{\mu} - m) \psi \eq 0
\ee
corresponds to the representation of type I. Nevertheless Dirac
theory allows one to add further details in a description of
elementary particles. First to mention is the prediction of
antiparticles, which is expressed in the fact that $\psi$ ist not a
two but four component function. In the following we use the
representation
\[
 \gamma^{0} \eq \left( \begin{array}{cc}  0 & \id \\ \id & 0
 \end{array} \right) \: , \: \gamma^{i} \eq \left( \begin{array}{cc}
 0 & \sigma_{i} \\ -\sigma_{i} & 0 \end{array} \right)  \, , \,
 i=1,2,3
\]
and
\[
 \gamma_{5} \eq \ii \gamma^{0} \gamma^{1} \gamma^{2} \gamma^{3}
\]
of the Dirac matrices. Then the notation of the reflections in the
Dirac theory of type I leads to:
\[ \begin{array}{ll}
 \mbox{space inversion: } & \psi (x,t) \mapsto \gamma^{0}
 \psi (-x,t) \: =: \: \psi_{{\rm s}} (x,t) \\
 \mbox{time inversion\footnotemark[2]: } & \psi (x,t) \mapsto
 \gamma^{1} \gamma^{3} {\rm J} \psi (x,-t)  \: =: \: \psi_{{\rm t}}
 (x,t) \\
 \mbox{charge conjug.\footnotemark[2]: } & \psi (x,t) \mapsto \ii
 \gamma^{2} {\rm J} \psi (x,t) \: =: \: \psi_{{\rm c}} (x,t)
\end{array} \]
Important for our aim to describe an electric dipole moment of an
elementary particle is to implement an interaction with an external
electromagnetic field. This is done by minimal coupling:
\be \label{d2}
 (\gamma^{\mu} (\ii \partial_{\mu} + A_{\mu}) - m)\psi \eq 0
\ee
Here $A = (A_{\mu})$ is a four-potential, which should transform
under the reflections as
\[ \begin{array}{ll}
 \mbox{space inversion: } & A(x,t) \mapsto \id_{1,3} A(-x,t) \\
 \mbox{time inversion: } & A(x,t) \mapsto \id_{1,3} A(x,-t) \\
 \mbox{charge conjugation: } & A(x,t) \mapsto -A(x,t)
\end{array} \]
To every four-potential there is an electromagnetic field tensor
$F_{\mu \nu} (x,t) \eq \partial_{\mu} A_{\nu} (x,t) - \partial_{\nu}
A_{\mu} (x,t)$. Its behaviour under reflections follows from the
behaviour of the corresponding four-potential. \\
space inversion:
\[
 F_{\mu \nu} (x,t) \mapsto \left\{
 \begin{array}{r@{\quad:\quad}l} F_{\mu \nu} (-x,t) & \mu ,\nu \eq
 1,\ldots ,3 \\ -F_{\mu \nu} (-x,t) & \mbox{else} \end{array}
 \right. \\
\]
time inversion:
\[
 F_{\mu \nu} (x,t) \mapsto \left\{
 \begin{array}{r@{\quad:\quad}l} -F_{\mu \nu} (x,-t) & \mu ,\nu \eq
 1,\ldots ,3 \\ F_{\mu \nu} (x,-t) & \mbox{else} \end{array}
 \right.
\]
charge conjugation:
\[
 F_{\mu \nu} (x,t) \mapsto -F_{\mu \nu} (x,t)
\]
Further it is important that the Dirac equation can be derived from
a Lagrangian. The Lagrangian
\be \label{l1}
 {\cal L} \eq \psi^{+} (\gamma^{\mu} (\ii \partial_{\mu} +e A_{\mu})
 - m)\psi \: , \: \psi^{+} \eq \gamma^{0} \psi^{*}
\ee
leads to equation (\ref{d2}), for which the following statement holds:

\begin{note}
Let $\psi$ be a solution of equation (\ref{d2}) then
$\psi_{{\rm s}}$ and $\psi_{{\rm t}}$ are solutions of the reflected
equations. These are the equations one gets by replacing ($x,t$) by
($-x,t$) or ($x,-t$) and the external fields by the reflected ones.
\end{note}

{\it Proof:} Follows immediatly from the behaviour of the
four-potential under the reflections and the commutation relations
of the gamma matrices.

The concept of a Lagrangian allows one to add effective terms which
are covariant and gauge invariant. So the addition of
\be
 {\cal L}_{{\rm eff}} \eq -d \psi^{+} \gamma^{\mu} \gamma^{\nu}
 \gamma_{5} F_{\mu \nu} \psi
\ee
to the Lagrangian (\ref{l1}) leads to the equation
\be \label{d3}
 (\gamma^{\mu} (\ii \partial_{\mu} + eA_{\mu}) - d\gamma^{\mu}
 \gamma^{\nu} \gamma_{5} F_{\mu \nu} - m)\psi \eq 0 \: .
\ee
An elementary particle with an electric dipol moment is described by
this equation. In the non-relativistic limit the coupling with
an external electric field is of the form
$\hat{\bf \sigma}{\bf E}$. Here is
\be \label{sghelp}
 \hat{\bf \sigma} = (\hat{\sigma_{1}} ,\hat{\sigma_{2}},
 \hat{\sigma_{3}} ) \: , \:
 \hat{\sigma_{i}} = \left( \begin{array}{cc} \sigma_{i} & 0 \\ 0 &
 \sigma_{i} \end{array} \right)
\ee
with ordinary Pauli matrices $\sigma_{i}$. As a consequence of the
added term one finds:

\begin{note}
Let $\psi$ be a solution of equation (\ref{d3}) then
$\psi_{s}$ and $\psi_{t}$ are no longer solutions of the reflected
equations.
\end{note}

This phenomenon is called violation of invariance under space
respectively time inversion. The situation is dif\-ferent when
looking at representations of type III. First one has to expand the
Dirac theory to representations with doubling. This is done for the
type III case by
\be \label{dd}
 (\ii \Gamma^{\mu} \partial_{\mu} - m)\psi \eq 0 \: , \: \Gamma^{\mu}
 \eq \left( \begin{array}{cc} \gamma^{\mu} & 0 \\ 0 & \gamma^{\mu}
 \end{array} \right) \: .
\ee
Here the notation of the reflections leads to:
\[ \begin{array}{ll}
 \mbox{space inversion: } & \psi (x,t) \mapsto \left(
                            \begin{array}{cc} \gamma^{0} & 0 \\
                            0 & -\gamma^{0} \end{array} \right)
                            \: =: \: \psi_{{\rm s}} (x,t) \\
 \mbox{time inversion: } & \psi (x,t) \mapsto \left(
                           \begin{array}{cc} 0 & \gamma^{1}
                           \gamma^{3} \\ -\gamma^{1} \gamma^{3} & 0
                           \end{array}  \right) {\rm J}
                           \psi (x,-t) \: =: \: \psi_{{\rm t}}
                           (x,t) \\
 \mbox{charge conjug.: } & \psi (x,t) \mapsto \ii \left(
                                \begin{array}{cc} \gamma^{2} & 0 \\
                                0 & \gamma^{2} \end{array} \right)
                                {\rm J} \psi (x,t) \: =: \:
                                \psi_{{\rm c}} (x,t)
    (x,t)
\end{array} \]
Now $\psi$ is an eight component function. An elementary particle
described by equation (\ref{dd}) has the same properties as a
particle described by equation (\ref{d1}), i.e.\ the same magnetic
dipole moment with the same behaviour under the reflections. Again
one can get the Dirac equation from a Lagrangian. Adding the
covariant and gauge invariant term
\be
 {\cal L}_{{\rm eff}} \eq -d \, \psi^{+} \left( \begin{array}{cc}
 0 &  \gamma^{\mu} \gamma^{\nu} \gamma_{5} \\ \gamma^{\mu}
 \gamma^{\nu} \gamma_{5} & 0 \end{array} \right) F_{\mu \nu} \psi
\ee
to the Lagrangian
\be
 {\cal L} \eq \psi^{+} (\Gamma^{\mu} (\ii \partial_{\mu} + eA_{\mu})
 - m) \psi
\ee
leads to the field equation
\begin{eqnarray} \label{d4}
 \left( \Gamma^{\mu} (\ii \partial_{\mu} + eA_{\mu}) -
 d\left(\begin{array}{cc} 0 & \gamma^{\mu} \gamma^{\nu} \gamma_{5} \\
 \gamma^{\mu} \gamma^{\nu} \gamma_{5} & 0 \end{array} \right)
 F_{\mu \nu} - m \right) \psi \eq 0 \: .
\end{eqnarray}
This equation again describes an elementary particle with
an electric dipole moment. In the non-relativistic limit the
coupling with an external electric field is of the form
\[
 \left( \begin{array}{cc} 0 & \hat{\bf \sigma} \\
 \hat{\bf \sigma} & 0 \end{array} \right) {\bf E} \: ,
\]
with $\hat{\bf \sigma}$ from (\ref{sghelp}). In contrast now to
the non-doubled case one has:

\begin{note}
Let $\psi$ be a solution of equation (\ref{d4}) then
$\psi_{s}$ and $\psi_{t}$ are solutions of the reflected equations.
\end{note}

This means that equation (\ref{d4}) is invariant both under space
and time inversion. Since equations (\ref{d3}) and (\ref{d4}) are
invariant under charge conjugation, which is seen by using the
commutation relations of the gamma matrices, the second equation
describes contrary to the first one a non-CP-violating electric
dipole moment of elementary particles.

\section*{Quantization of the Dirac Field}

We write the field equation (\ref{d4}) in terms of a Hamiltonian
which can be checked to be hermitian. That means:
\be
 \ii \partial_{t} \psi \eq H\psi
\ee
with
\be
 H = \Gamma^{0} \Gamma^{j} (\ii \partial_{j} - eA_{j}) + eA_{0}
     +\Gamma^{0} m \nonumber -d\, \Gamma^{0} \left(
     \begin{array}{cc} 0 & \gamma^{\mu} \gamma^{\nu} \gamma_{5}
     \\ \gamma^{\mu} \gamma^{\nu} \gamma_{5} \end{array} \right)
     F_{\mu \nu} \nonumber
\ee
The spectrum of the Hamiltonian $H$ depends on the external
fields. We want to assume that the Hilbert space ${\cal H}$ of
the Dirac equation can be split into two orthogonal spectral
subspaces of the Hamiltonian,
\[
 {\cal H} \eq {\cal H}_{+} \oplus {\cal H}_{-} \: ,
\]
such that ${\cal H}_{+}$ can be interpreted as a Hilbert space for a
particle. This is the assumption of weak external fields. For a more
rigorous treatment see \cite{thaller}.

Now one is able to build the Fock space ${\cal F}$ over ${\cal H}$.
The unitary or antiunitary operators belonging to Poincar\'e
transformations including the reflections can be implemented by
unitary and antiunitary operators, respectively, because these
operators do not mix the Hilbert spaces ${\cal H}_{+}$ and
${\cal H}_{-}$. See \cite{thaller} again for further details.

To complete our discussion we want to give the transformation
formulas of the Dirac operator $\psi (x)$ under the reflections in
the type III representation. Therefore we write the Dirac operator as
\begin{eqnarray}
\psi (x) & = & \frac{1}{(2\pi)^{3/2}} \int
               \frac{{\rm d}^{3}\vec{p}}{2\omega_{\vec{p}}} \bigg(
               \exp (\ii( \omega_{\vec{p}} x^{0} - \vec{p} \vec{x}))
               \sum_{r,s=1}^{2} v^{rs} (\vec{p}) {\bf b}_{rs}^{+}
               (\vec{p}) \nonumber \\
         &   & + \exp (-\ii (\omega_{\vec{p}} x^{0} - \vec{p} \vec{x}))
               \sum_{r,s=1}^{2} u^{rs} (\vec{p}) {\bf a}_{rs}
               (\vec{p}) \bigg)
\end{eqnarray}
with
\[
 v^{1s} (\vec{p}) \eq \left( \begin{array}{c} \hat{v}^{s} (\vec{p})
 \\ 0 \end{array} \right) \: , \: v^{2s} (\vec{p}) \eq \left(
 \begin{array}{c} 0 \\ \hat{v}^{s} (\vec{p}) \end{array} \right)
\]
\[
 u^{1s} (\vec{p}) \eq \left( \begin{array}{c} \hat{u}^{s} (\vec{p})
 \\ 0 \end{array} \right) \: , \: u^{2s} (\vec{p}) \eq \left(
 \begin{array}{c} 0 \\ \hat{u}^{s} (\vec{p}) \end{array} \right)
\]
where $\hat{v}^{s} (\vec{p})$ and $\hat{u}^{s} (\vec{p})$ are
solutions of the Fourier-transformed field equation (\ref{d1}). The
unitary operators $U_{{\rm p}}, U_{{\rm c}}$ and the antiunitary
operator $U_{{\rm t}}$ are specified by
\begin{eqnarray}
 U_{{\rm p}}\psi (x)U^{-1}_{{\rm p}} & = & \left( \begin{array}{cc}
 \gamma^{0} & 0 \\ 0 & -\gamma^{0} \end{array} \right) \psi (-x,t)
 \nonumber \\
 U_{{\rm t}}\psi (x)U^{-1}_{{\rm t}} & = & \left( \begin{array}{cc}
 0 & \gamma^{1} \gamma^{3} \\ -\gamma^{1} \gamma^{3} \end{array}
 \right) \psi (x,-t) \nonumber \\
 U_{{\rm c}}\psi (x)U^{-1}_{{\rm c}} & = & \ii \left(
 \begin{array}{cc} \gamma^{2} &  0 \\ 0 & \gamma^{2} \end{array}
 \right) {\rm J} \psi (x,t) \: .
\end{eqnarray}
The field equations for the Dirac operator $\psi (x)$ remain
invariant under the reflections. A little calculation which is
similar to the case of the type I representation leads to
\begin{eqnarray}
 U_{{\rm p}} {\rm a}_{rs}^{+} U_{{\rm p}}^{-1} & = &
 {\rm a}_{rs}^{+} (-\vec{p}) \nonumber \\
 U_{{\rm p}} {\rm b}_{rs}^{+} U_{{\rm p}}^{-1} & = &
 -{\rm b}_{rs}^{+} (-\vec{p})  \nonumber \\
 U_{{\rm t}} {\rm a}_{rs}^{+} U_{{\rm t}}^{-1} & = &
 {\rm a}_{r'(-s)}^{+}  (-\vec{p}) \nonumber \\
 U_{{\rm t}} {\rm b}_{rs}^{+} U_{{\rm t}}^{-1} & = &
 {\rm b}_{r'(-s)}^{+} (-\vec{p}) \nonumber \\
 U_{{\rm c}} {\rm a}_{rs}^{+} U_{{\rm c}}^{-1} & = &
 {\rm b}_{rs}^{+} (\vec{p}) \nonumber \\
 U_{{\rm c}} {\rm b}_{rs}^{+} U_{{\rm c}}^{-1} & = &
 {\rm a}_{rs}^{+} (\vec{p})
\end{eqnarray}
If we require that these equations leave the vacuum $\Omega \in
{\cal F}$ invariant,
\be
 U_{{\rm p}} \Omega \eq U_{{\rm t}} \Omega \eq U_{{\rm c}} \Omega
 \eq \Omega \: ,
\ee
then $U_{{\rm p}}, U_{{\rm t}}$ and $U_{{\rm c}}$ are again unitary
or antiunitary operators in the Fock space.

\section*{Acknowledgement}
I thank H.\ Reeh, who proposed to look at the representations
of the Poincar\'e group.

\end{document}